\documentclass[twocolumn,showpacs,preprintnumbers,amsmath,amssymb]{revtex4}

\usepackage{graphicx}
\usepackage{dcolumn}
\usepackage{bm}

\begin{document}

\title{Quasinormal Modes of Black Holes and Dissipative Open
Systems}

\author{Sang Pyo Kim}\email{sangkim@kunsan.ac.kr}

\affiliation{Department of Physics, Kunsan National University,
Kunsan 573-701, Korea}

\affiliation{Asia Pacific Center for Theoretical Physics, Pohang
790-784, Korea}

\date{\today}
\begin{abstract}
After explaining the physical origin of the quasinormal modes of
perturbations in the background geometry of a black hole, I
critically review the recent proposal for the quantization of the
black-hole area based on the real part of quasinormal modes. As
instantons due to the barriers of black-hole potentials lie at the
root of a discrete set of complex quasinormal modes frequencies,
it is likely that the physics of quasinormal modes can be learned
from quantum theory. I propose a connection of a system of
quasinormal modes of black holes with a dissipative open system,
in particular, the Feshbach-Tikochinsky oscillator. This argument
is supported in part by the fact that these two systems have the
same group structure $SU(1,1)$ and the same group representation
of Hamiltonians; thereby, their quantum states exhibit the same
behavior. 
\end{abstract}

\maketitle

\section{Introduction}

In linear perturbation theory, a perturbed black hole emits waves
that are outgoing to spatial infinity and the event horizon. The
wave function with this boundary condition is called a quasinormal
mode of the black hole and has a complex frequency, whose
imaginary part leads to a decaying behavior. There are several
motivations to study quasinormal modes of black holes. The
original motivation is that quasinormal modes carry carry
information of black hole, such as mass, charge, and angular
momentum, and may be observed by a gravitational wave detector
\cite{vishveshwara}. Another motivation comes from a recent
argument that the real part of the quasinormal mode frequency may
play a role in explaining the quantization of the black-hole area
\cite{hod,kunstatter,horowitz}. The quantization of the black-hole
area, which is believed to be closely related with the microscopic
origin of black-hole entropy, is also explained in loop quantum
gravity \cite{dreyer}. Ideas have been suggested to understand the
origin of quasinormal modes and various methods have been
developed to find quasinormal modes (for review and references,
see Refs. \cite{kokkotas,natario}). Quite recently, I have
suggested that the quantum theory of quasinormal modes might be
related with a dissipative open system and possibly with
black-hole thermodynamics \cite{kim0,kim1}.

The purpose of this paper is {\it not} to develop any new method
for finding analytically quasinormal modes, {\it but} to exploit
the physical interpretation of quasinormal modes of black holes.
In particular, I shall provide some supporting arguments for my
recent proposal on the connection of quasinormal modes with a
dissipative open system, the Feshbach-Tikochinsky (FT) oscillator,
and possibly with the thermodynamics of black holes in a
semiclassical approach. There are some open questions about
quasinormal modes and thermodynamics. First, is there any
connection between a system of quasinormal modes and a dissipative
open system? Second, is there any connection between quasinormal
modes and black-hole thermodynamics? It is known that
perturbations of a system may provide some information of the
system, and, in some cases, may carry all information. Then, what
black-hole physics can we learn from quasinormal modes?

\section{What are Quasinormal Modes of Black Holes?}

Let us consider perturbations in a Schwarzchild black hole with
mass $M$. The equation for perturbations is given by
\begin{equation}
\frac{\partial^2}{\partial t^2} \Psi - \frac{\partial^2}{\partial r_*^2}
+ V(r_*) \Psi = 0,
\end{equation}
where $r_* = r + 2M \ln(r/2M -1)$ is the tortoise coordinate. In
the mode decomposition
\begin{equation}
\Psi_{\omega l m} (t, r, \theta, \phi) = e^{i \omega t}
\psi_{(\omega l)} (r_*) Y_{lm} (\theta, \phi), \label{qnm mod}
\end{equation}
the perturbation equation is separated as
\begin{equation}
- \frac{\partial^2}{\partial r_*^2} \psi_{(\omega l)} + V_{(l)}
(r_*) \psi_{(\omega l)} = \omega^2 \psi_{(\omega l)}.
\end{equation}
The potential of the Schwarzschild black hole is given by
\begin{equation}
V_{(l)} = \Bigl(1 - \frac{2M}{r} \Bigr) \Bigl( \frac{l(l+1)}{r^2}
+ \frac{2 \sigma M}{r^3} \Bigr),
\end{equation}
where scalar $(l \geq 0)$, electromagnetic $(l \geq 1)$ and
gravitational $(l \geq 2)$ perturbations have $\sigma = 1, 0, -3$,
respectively. For instance, Fig. 1 shows the first few black hole
potentials for the scalar perturbation. There is a potential
barrier for each $l$.
\begin{figure}
\includegraphics[scale=0.75]{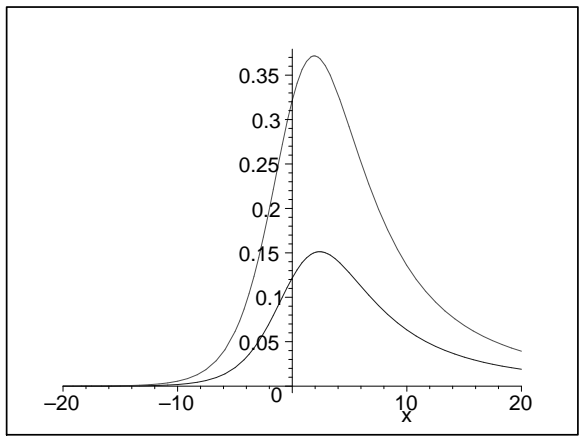}
\caption{The black hole potentials for scalar perturbation are
drawn in the scale of black hole mass $M =1$ for $l = 2$ (lower
graph) and $l =3$ (upper graph).}
\end{figure}

To obtain the quasinormal mode of the black hole, one imposes a
boundary condition such that the mode is an outgoing wave to the
spatial infinity and the event horizon. Though many numerical
methods have been developed to find quasinormal-mode frequencies
(for instance, see the references in Refs.
\cite{kokkotas,natario}), exact analytical solutions have not yet
been found.

Physically, quasinormal modes can be interpreted as a scattering
problem as follows: An incident wave with a unit amplitude is
partly reflected with an amplitude $R$ and partly transmitted
through the barrier with an amplitude $T$ at asymptotical regions
in the tortoise coordinate. The branches of the wave for the
scattering problem can be written as
\begin{eqnarray}
\Psi_{in} \approx \frac{1}{R(\omega)} e^{i \omega (t + r_*)},
\quad \Psi_{re} \approx e^{i \omega (t - r_*)},
\end{eqnarray}
and for the transmitted wave as
\begin{eqnarray}
\Psi_{tr} \approx \frac{T(\omega)}{R(\omega)} e^{i \omega (t +
r_*)}.
\end{eqnarray}
To have a purely outgoing wave to the spatial infinity and the
event horizon, $R$ should be singular to make the incident wave
negligible compared with the reflected wave, but $T/R$ should be
regular to have a finite transmission amplitude. This means that
$T$ is also singular. The consequence of the boundary condition
will be discussed in Sec. IV.

\section{Quasinormal Modes and Black Hole Area Quantization}

The quantization of the black hole area was first proposed by
Bekenstein \cite{bekenstein}. Recently, it has been proposed that
the real parts of quasinormal modes may explain the quantization
of the black-hole area. Hod first related the real parts of
quasinormal modes with the quantization of the black-hole area by
using Bohr's correspondence principle \cite{hod}. The
correspondence principle is a semiclassical idea that wave
functions with large quantum numbers or actions follow almost
classical orbits and exhibit classical behavior and that classical
theory can be obtained from quantum theory in this limit.

Quasinormal-mode frequencies near the top of the black-hole
potential have an asymptotic form \cite{nollert}
\begin{equation}
M \omega_n = 0.0437123 + i \frac{1}{4} \Bigl(n + \frac{1}{2}
\Bigr) + {\cal O} \Bigl( \frac{1}{(n+ 1/2)^{1/2}}\Bigr).
\label{qnm freq}
\end{equation}
The real parts of quasinormal modes are responsible for the
oscillatory behavior and, thus, play the role of the quantum
action in the correspondence principle whereas the imaginary parts
are responsible for the decaying behavior. Note that $M \omega_R
\simeq \ln 3/(8 \pi)$. As the mass of the black hole changes by
the minimum energy quantum, $dM = E = \hbar \omega_R$, the area,
$A = 16 \pi M^2$, of the event horizon changes by $dA = 4 \ln 3
\times l_p^2$. For a Kerr black hole, one has $\omega_R = \ln 3
\times T_{bh} + \Omega m$ \cite{hod2}. Therefore, it seems that
Hod's argument explains the quantization of the black-hole area
from a knowledge of quasinormal modes.

Kunstatter further noted that the real parts of quasinormal modes
are given by the transit time of light across the horizon
\cite{kunstatter}:
\begin{equation}
\omega_R = \alpha^{(4)} \Bigl( \frac{c}{R_H} \Bigr),
\end{equation}
where $2(R_H = 2M)/c$ is the transit time across the horizon and
$\alpha^{(4)} = 0.0437123 \times 2 = \ln 3/(4 \pi)$. Then,
according to the Bohr-Sommerfeld quantization rule, the adiabatic
invariant is related with the entropy of the black hole as
\begin{equation}
I = \int \frac{dE}{\omega_R} = \frac{c^3}{2 G \alpha^{(4)}} \int E
dE = \frac{\hbar}{4 \pi \alpha^{(4)}} S_{bh} = n \hbar.
\end{equation}
Thus, the black-hole entropy is given by $S_{bh} = 4 \pi
\alpha^{(4)} n = \ln 3 \times n$, and the number of states by
$\Omega(E) = e^{S_{bh}} = 3^n$.

In the above argument for the quantization of the black-hole area,
the real parts of quasinormal modes provide the minimum energy for
excitation of the black-hole area. Quasinormal modes have a
quantum mechanical analog: a particle confined to a potential well
by a finite potential barrier has a complex energy whose real part
is responsible for a quasi-stationary state inside the potential
well and whose imaginary part gives the decay rate. Both the real
and the imaginary parts have physical meanings in quantum theory.
However, the above argument used only the real parts of
quasinormal modes.

\section{Origin of Discrete Imaginary Frequencies}

Black-hole potentials in a four-dimensional spacetime are too
complicated to allow any analytical method, so one stratagem is to
approximate the black-hole potentials by some exactly solved ones.
Certain kinds of simulated potentials are known to possess group
structures which yield the spectrum through a group theoretical
technique. The physical origin of the discrete set of imaginary
parts may be illustrated by an inverted potential model, which
approximates the top region of the black hole potential.

The inverted potential
\begin{equation}
\tilde{H} = \frac{p^2}{2} - \frac{\omega_0^2}{2} x^2 + V_0
\end{equation}
is obtained by analytically continuing the frequency $\omega_0
\rightarrow i \omega_0$ of a harmonic oscillator
\begin{equation}
H = \frac{p^2}{2} + \frac{\omega_0^2}{2} x^2 + V_0.
\end{equation}
In terms of the annihilation and creation operators
\begin{eqnarray}
\hat{a} = \frac{1}{\sqrt{2 \omega_0}} (\hat{p} - i \omega_0
\hat{x}), \quad \hat{a}^{\dagger} = \frac{1}{\sqrt{2
\omega_0}} (\hat{p} + i \omega_0 \hat{x}),
\end{eqnarray}
the harmonic oscillator has the representation
\begin{equation}
\hat{H} = \omega_0 \Bigl(\hat{a}^{\dagger} \hat{a} + \frac{1}{2}
\Bigr) + V_0.
\end{equation}
The oscillator has the group $SU(1,1)$ consisting of generators
$\hat{a}^2$, $\hat{a}^{\dagger 2}$, and $\hat{a}^{\dagger}
\hat{a}$. The group generates all energy states from a given
energy state $\hat{H} \vert \epsilon \rangle = \epsilon \vert
\epsilon \rangle$ as
\begin{eqnarray}
\hat{H} \hat{a}^{\dagger} \vert \epsilon \rangle = ( \epsilon +
\omega_0) \hat{a}^{\dagger} \vert \epsilon \rangle, \quad
\hat{H} \hat{a} \vert \epsilon \rangle =
(\epsilon - \omega_0) \hat{a} \vert \epsilon \rangle,
\end{eqnarray}
from which follows the energy spectrum
\begin{equation}
 \frac{{\cal E}_n}{\omega_0} = \frac{V_0}{\omega_0} + \Bigl(n +\frac{1}{2} \Bigr).
\end{equation}

Similarly, the inverted oscillator has the representation for the
annihilation and creation operators \cite{barton}
\begin{eqnarray}
\hat{b} = \frac{e^{ i \theta}}{\sqrt{2 \omega_0}} (\hat{p} +
\omega_0 \hat{x}), \quad \hat{b}^{\dagger} = \frac{e^{- i
\theta}}{\sqrt{2 \omega_0}} (\hat{p} - \omega_0 \hat{x}),
\end{eqnarray}
for a real $\theta$. Note that the position is antiunitary,
$\hat{x}^{\dagger} = - \hat{x}$, whereas the momentum is
hermitian:
\begin{eqnarray}
\hat{x} &=& \frac{1}{\sqrt{2 \omega_0}} (e^{-i \theta} \hat{b} -
e^{i \theta} \hat{b}^{\dagger}), \nonumber\\
\hat{p} &=& \sqrt{\frac{\omega_0}{2}} (e^{-i \theta} \hat{b} +
e^{i \theta} \hat{b}^{\dagger}).
\end{eqnarray}
The annihilation and the creation operators satisfy the
commutation relation
\begin{equation}
[ \hat{b}, \hat{b}^{\dagger}] = i.
\end{equation}
The Hamiltonian has the representation
\begin{equation}
\hat{\tilde{H}} = \omega_0 \Bigl(\hat{b}^{\dagger} \hat{b} +
\frac{i}{2} \Bigr) + V_0.
\end{equation}
Thus, all energy eigenstates are constructed from a given energy
state $\hat{\tilde H} \vert \epsilon \rangle = \epsilon \vert
\epsilon \rangle$ as
\begin{eqnarray}
\hat{\tilde H} \hat{b}^{\dagger} \vert \epsilon \rangle = (
\epsilon + i \omega_0) \hat{b}^{\dagger} \vert \epsilon \rangle,
\quad \hat{\tilde H} \hat{b} \vert \epsilon \rangle = (
\epsilon - i \omega_0) \hat{b} \vert \epsilon \rangle,
\end{eqnarray}
The corresponding spectrum has quantized imaginary values
\begin{equation}
\frac{\tilde{\cal E}_n}{\omega_0} = \frac{V_0}{\omega_0} + i \Bigl( n +
\frac{1}{2} \Bigr). \label{inv en}
\end{equation}

The wave function for the inverted oscillator is given by
\begin{equation}
\Psi (x) = E(s, \sqrt{2 \omega_0}x),
\end{equation}
where $E$ is a complex parabolic cylindrical function and $s =
(\tilde{\cal E} - V_0)/\omega_0$. The scattering matrix, the ratio
of the reflected amplitude to the incident amplitude, is
\begin{equation}
{\cal M} = - \Bigl(\frac{1}{1 + e^{- 2 \pi s}} \Bigr)^{1/2}.
\end{equation}
Thus, the condition for quasinormal modes given by
\begin{equation}
\pi s = \pi \frac{\tilde{\cal E} -V_0}{\omega_0} = i \pi \Bigl(n +
\frac{1}{2} \Bigr)
\end{equation}
is that instanton actions should be quantized as $S_n = \pi s$
\cite{kim2}. Note that the quasinormal-mode frequencies of a
black-hole potential and the energy spectrum of the inverted
oscillator have the same discrete set of imaginary parts up to
multiplication factors. The high-toned quasinormal modes come from
a region near the top of black hole potential, which is well
approximated by the inverted oscillator.

Another exactly solved potential is provided by using a
self-similar collapsing scalar field model
\cite{roberts,bak1,bak2,bak3}. The spherically symmetric geometry
minimally coupled to a massless scalar field is given by the
metric
\begin{equation}
ds^2 = - 2 du dv  + r^2 d\Omega_2^2.
\end{equation}
For a self-similar collapse, we use the coordinates
\begin{equation}
r = \sqrt{- uv} y(z), \quad z = - \frac{v}{u} = e^{-2 \tau},
\end{equation}
or
\begin{equation}
u = - \omega e^{- \tau}, \quad v =  \omega e^{\tau}.
\end{equation}
The quantum gravitational collapse is then described by the
Wheeler-DeWitt equation \cite{bak2,bak3}
\begin{equation}
\Biggl[\frac{1}{2K} \frac{\partial^2}{\partial y^2} -
\frac{m_P^2}{2K y^2} \frac{\partial^2}{\partial \phi^2} - K \Bigl(
1 - \frac{y^2}{2}\Bigr) \Biggr] \Psi (y, \phi)= 0,
\end{equation}
where $K = m_P^2 \omega_c^2/2$. If the wave function is separated
as $\Psi = e^{ \pm i (Kc_0 / m_P) \phi } \psi$, the Wheeler-DeWitt
equation now takes the form
 \begin{equation}
\Biggl[- \frac{1}{2K} \frac{\partial^2}{\partial y^2} +
\frac{K}{2} \Biggl(2  - y^2 -  \frac{c_0^2}{y^2}\Bigr) \Biggr]
\psi (y)= 0. \label{sol mod}
\end{equation}
Equation (\ref{sol mod}) is the Schr\"{o}dinger equation for a
particle with mass $K$ and constant energy in an inverted
oscillator potential, but with an additional inverse-squared term.

The wave equation, (\ref{sol mod}), can also be solved
analytically in terms of the confluent hypergeometric function or
group theoretically by using $SU(1,1)$. In fact, the
inverse-squared term does not change the group. There are three
classes of solutions: supercritical collapse for $c_0 > 1$,
critical collapse for $c_0 = 1$, and subcritical collapse for $c_0
< 1$. The case relevant to quasinormal modes is subcritical
collapse, which describes the quantum-mechanical formation of
black holes. The boundary condition for black hole formation that
the wave has an outgoing flux to spatial infinity and the event
horizon is the same as that for quasinormal modes. Such a wave
function is found to be \cite{bak2}
\begin{equation}
\Psi_{bh} = e^{ - iK y^2/2} (Ky^2)^{\mu} M (a, b, iKy^2),
\end{equation}
where $M$ is the confluent hypergeometric function and
\begin{eqnarray}
a &=& \frac{1}{2} - \frac{i}{2} (Q+K), \nonumber\\
b &=& 1 - i Q, \nonumber\\
\mu &=& \frac{1}{4} - \frac{i}{2} Q, \nonumber\\ Q &=&
\Bigl(K^2c_0^2 - \frac{1}{4} \Bigr)^{1/2}.
\end{eqnarray}
The scattering matrix, the ratio of the reflected amplitude to the
incident amplitude at spatial infinity, is given by
\begin{equation}
{\cal M} = \frac{\Gamma(b - a)}{\Gamma(a)} (iK)^{2a - b} e^{- i
\pi}.
\end{equation}
The scattering matrix has simple poles at
\begin{equation}
b - a = - \frac{i}{2} (Q - K) + \frac{1}{2} = - n, \quad (n = 0,
1, \cdots),
\end{equation}
and, thus, restricts $c_0$ to
\begin{equation}
\frac{1}{2} \Bigl[ K - \Bigl(K^2c_0^2 - \frac{1}{4} \Bigr)^{1/2}
\Bigr] = i \Bigl(n + \frac{1}{2} \Bigr).
\end{equation}
The discrete set of imaginary values is a consequence
of $SU(1,1)$ of the model potential.

The black-hole potential can also be approximated by a
P\"{o}schl-Teller potential \cite{ferrai}
\begin{equation}
V_{PT}(x) = \frac{V_0}{\cosh^2 (\alpha (x- x_0))},
\end{equation}
where $V_0$, $w$, and $x_0$ denote the height, the inverse of the
width and the center of the potential, respectively. The wave
functions for quasinormal modes can be found in terms of the
hypergeometric function \cite{ferrai2}, which leads to the complex
frequencies
\begin{equation}
\omega = \Bigl( V_0 - \frac{\alpha^2}{4} \Bigr)^{1/2} + i \alpha
\Bigl( n + \frac{1}{2} \Bigr), \quad (n = 0, 1, 2, \cdots).
\end{equation}
Here, the imaginary parts can be interpreted as instantons. The
solvability and the set of discrete values of the imaginary parts
are a consequence of the group $SU(2)$ of the P\"{o}schl-Teller
potential. To improve the simulated potential and take into
account the asymmetry of the black-hole potential, one may use the
generalized P\"{o}schl-Teller potential
\begin{equation}
V_{GPT}(x) = \frac{V_0 + U_1 \sinh_q (\alpha (x - x_0))}{\cosh^2_q
(\alpha (x- x_0))},
\end{equation}
where $\cosh_q x = (e^x + qe^{-x})/2$ and $\sinh_q x = (e^x - q
e^{-x})/2$ are q-deformed functions of $\cosh x$ and $\sinh x$.
The generalized P\"{o}schl-Teller potential also has the group
$SU(2)$ and thereby the complex frequencies
\begin{equation}
\omega = \frac{1}{\sqrt{2}} \Biggl( V_0 - \frac{\alpha^2}{4}
 + \Bigl((V_0 - \frac{\alpha^2}{4})^2
 + \frac{V_1^2}{q} \Bigr)^{1/2} \Biggr)^{1/2}
 + i \alpha \Bigl( n + \frac{1}{2}
 \Bigr).
\end{equation}
The scattering matrix has also been calculated using the singular
structure of the black-hole potential \cite{motl,motl2}.

\section{Second Quantization of Quasinormal Modes}

The quantization of unstable systems has been one of the issues in
quantum theory. For instance, the scalar field (tachyon) with a
negative mass was studied a long time ago \cite{schroer,schroer2}.
The Fourier-decomposed modes have both growing and decaying
solutions. Another system is provided by a scalar field in the
Kerr background geometry with or without a horizon, and the
quantization of unstable modes has been treated in Refs.
\cite{matacz,kang,mukohyama}. In contrast with the instability of
a field in the rotating black hole geometry, quasinormal modes are
stable in the sense of decaying in time. The stability of
quasinormal modes comes from the boundary condition.

To quantize a scalar field in a curved spacetime, one has to
introduce an inner product and thereby a Hilbert space for the
field \cite{fulling}. The scalar field
has the well-known Klein-Gordon norm
\begin{equation}
(\Psi_{\lambda}, \Psi_{\lambda'}) = - i \int (\Psi_{\lambda}^*
\overleftrightarrow{\partial_t} \Psi_{\lambda'}) \sqrt{g} d^3x.
\label{kg nor}
\end{equation}
Here, the negative sign in the Klein-Gordon norm is chosen to fit
the definition of quasinormal modes in Eq. (\ref{qnm mod}). In the
Minkowski spacetime, a massive scalar field has, in spherical
coordinates, the modes
\begin{equation}
\Psi_{\omega lm} (t, r, \theta, \phi) = \sqrt{\frac{k}{\pi}} e^{i
\omega t} j_l(kr) Y_{lm} (\theta, \phi), \label{sp mod}
\end{equation}
where $j_l$ is the spherical Bessel function, and $\omega$ and $k$
are related by
\begin{equation}
\omega^2 = k^2 + m^2.
\end{equation}
The modes in Eq. (\ref{sp mod}) satisfy the norms
\begin{eqnarray}
(\Psi_{\omega l m}, \Psi_{\omega'l'm'}) &=& \delta(\omega -
\omega')
\delta_{ll'} \delta_{mm'}, \nonumber\\
(\Psi^*_{\omega l m}, \Psi_{\omega'l'm'}) &=& 0, \nonumber\\
(\Psi^*_{\omega l m}, \Psi^*_{\omega'l'm'}) &=& - \delta(\omega -
\omega') \delta_{ll'} \delta_{mm'}.
\end{eqnarray}
Then, the quantum field has the expansion
\begin{equation}
\hat{\Psi} = \sum_{\lambda} (\Psi_{\lambda} \hat{a}_{\lambda} +
\Psi_{\lambda}^* \hat{a}_{\lambda}^{\dagger}),
\end{equation}
where $\lambda = (\omega lm)$. The Fock space operators satisfy
the commutation relations
\begin{equation}
[\hat{a}_{\lambda}, \hat{a}_{\lambda'}^{\dagger} ] =
\delta_{\lambda \lambda'}, \quad [\hat{a}_{\bar{\lambda}},
\hat{a}_{\bar{\lambda}'}^{\dagger} ] = \delta_{\bar{\lambda}
\bar{\lambda}'},
\end{equation}
and the canonical Hamiltonian is given by
\begin{eqnarray}
\hat{H} = \frac{i}{2} (\hat{\Psi}, \partial_t \hat{\Psi}) =
\frac{1}{2}\sum_{\lambda}  \omega_{\lambda} (\hat{a}_{\lambda}
\hat{a}^{\dagger}_{\lambda} + \hat{a}_{\lambda}^{\dagger}
\hat{a}_{\lambda}).
\end{eqnarray}

The scalar field in the background geometry of a Schwarzchild
black hole has not only static modes but also quasinormal modes,
depending on the imposed boundary condition. Hawking used static
modes with an appropriate boundary condition to find pair creation
and thereby Hawking radiation \cite{hawking}. From now on, I shall
focus on the quasinormal modes of the black hole. To quantize the
quasinormal modes, I shall adopt the stratagem taken by Mukohyama
for an unstable field \cite{mukohyama}. Our case of quasinormal
modes will be less difficult, because they are decaying and, thus,
stable. The inner product will be provided by the Klein-Gordon
norm in Eq. (\ref{kg nor}). As in the Minkowski spacetime where
there is a complete basis $\{\Psi_{\lambda}, \Psi^*_{\lambda} \}$,
the quasinormal modes given in Eq. (\ref{qnm mod}),
\begin{equation}
\Psi_{\lambda} = e^{ i \omega_{\lambda} t} \psi_{\lambda}, \quad
({\rm Im} (\omega_{\lambda}) > 0), \label{cl 1}
\end{equation}
have the conjugate modes
\begin{equation}
\Psi_{\bar\lambda} = e^{- i \omega^*_{\lambda} t}
\psi^*_{\lambda}, \quad ({\rm Im} (-\omega_{\lambda}) > 0).
\label{cl 2}
\end{equation}
The modes in the class in Eq. (\ref{cl 1}) are decaying in time
whereas those in the class in Eq. (\ref{cl 2}) are growing and,
thus, unstable. Further, the quasinormal modes $\Psi_{\lambda}$
have the boundary condition of outgoing to infinity and the
horizon whereas the modes $\Psi_{\bar\lambda}$ denote waves
incoming from infinity and the event horizon.

The Hilbert space, necessary in quantizing a field, can be
constructed with a complete basis. The complete basis is also
necessary to expand the field itself. Although the set
$\{\Psi_{\lambda}, \Psi_{\bar\lambda} \}$ in Eqs. (\ref{cl 1}) and
(\ref{cl 2}) has not been mathematically proved to form such a
complete basis, it is worth noting the proof by Beyer that the
P\"{o}schl-Teller potential, as an approximation of black hole
potentials in Sec. III, has a complete basis consisting of
quasinormal modes \cite{beyer}. He showed that after a large
enough time, any wave function with a compact support can be
expanded uniformly in quasinormal modes. Price and Husain also
proved the completeness of quasinormal modes in a model of
relativistic stellar oscillations \cite{price}. The completeness
of quasinormal modes was also intensively studied in the
one-dimensional wave equation with a certain boundary condition
and was used to quantize the quasinormal modes \cite{ching,ho}.

The completeness of the quasinormal modes of a black hole will be
assumed, and the proof will be deferred for a future work. That
is, $\{\Psi_{\lambda}, \Psi_{\bar\lambda} \}$ form a complete
basis. Then, a quantum field with a compact support can be
expanded as
\begin{equation}
\hat{\Psi} = \sum_{\lambda, \bar{\lambda}} (\Psi_{\lambda}
\hat{a}_{\lambda} + \Psi_{\lambda}^* \hat{a}_{\lambda}^{\dagger} +
\Psi_{\bar\lambda} \hat{a}_{\bar\lambda} + \Psi_{\bar\lambda}^*
\hat{a}_{\bar\lambda}^{\dagger}).
\end{equation}
Finally, the canonical Hamiltonian for quasinormal modes takes the
form
\begin{equation}
\hat{H}_{QNM} = \sum_{\lambda, \bar{\lambda}} \Bigl[ \omega_{R
\lambda} (\hat{a}^{\dagger}_{\lambda} \hat{a}_{\lambda} -
\hat{a}_{\bar{\lambda}}^{\dagger} \hat{a}_{\bar{\lambda}}) + i
\frac{\omega_{I \lambda}}{2} (\hat{a}^{\dagger}_{\lambda}
\hat{a}^{\dagger}_{\bar{\lambda}} - \hat{a}_{\lambda}
\hat{a}_{\bar{\lambda}}) \Bigr]. \label{qnm ham}
\end{equation}
The $\lambda$th mode Hamiltonian
\begin{equation}
\hat{H}_{\lambda} = 2 \omega_{R \lambda} \hat{\cal C} - \omega_{I
\lambda} \hat{J}_1,
\end{equation}
has an $SU(1,1)$ group in the Schwinger two-mode representation
\begin{eqnarray}
\hat{J}_1 &=& \frac{1}{2} (\hat{a}^{\dagger}_{\lambda}
\hat{a}^{\dagger}_{\bar\lambda} +
\hat{a}_{\lambda} \hat{a}_{\bar\lambda}), \nonumber\\
\hat{J}_2 &=& - \frac{i}{2} (\hat{a}^{\dagger}_{\lambda}
\hat{a}^{\dagger}_{\bar\lambda} - \hat{a}_{\lambda}
\hat{a}_{\bar\lambda}), \nonumber\\
\hat{J}_3 &=& \frac{1}{2} (\hat{a}^{\dagger}_{\lambda}
\hat{a}_{\lambda} + \hat{a}^{\dagger}_{\bar\lambda}
\hat{a}_{\bar\lambda} + 1), \label{su}
\end{eqnarray}
and the Casimir operator is given by
\begin{equation}
\hat{\cal C} = \frac{1}{2} (\hat{a}^{\dagger}_{\lambda}
\hat{a}_{\lambda} - \hat{a}_{\bar\lambda}^{\dagger}
\hat{a}_{\bar\lambda}).
\end{equation}
The group representation and quantum states will be given in the
next section.

\section{Connection with the Feshbach-Tikochinsky Oscillator}

A quasinormal mode is a decaying wave function due to its complex
frequency. In quantizing the system of quasinormal modes, it would
be of help and interest to compare it with other dissipative
systems. One well-known dissipative open system is the
Feshbach-Tikochinsky (FT) oscillator \cite{feshbach}. Feshbach and
Tikochinsky introduced a two-dimensional oscillator, with one
subsystem being damped and the other being amplified. The
subsystem of a damped oscillator is a physical system whereas the
subsystem of an amplified oscillator corresponds to an
environment. There is another damped oscillator model, an
oscillator with an exponentially growing mass, which is a unitary
theory \cite{kim3,kim4}. In this section, the quantum theory of
the FT oscillator will be reviewed and compared with the quantum
theory of quasinormal modes. I shall follow the development of
quantum theory in Ref. \cite{celeghini}.

The Lagrangian for the two-dimensional FT oscillator is given by
\begin{equation}
L_{FT} = \dot{x} \dot{y} + \frac{\gamma}{2} ( x \dot{y} - \dot{x}
y) - k xy.
\end{equation}
The $x$ coordinate describes the damped motion
\begin{equation}
\ddot{x} + \gamma \dot{x} + k x = 0.
\end{equation}
The damped solution is given by
\begin{eqnarray}
x = x_0 e^{i \omega t},
\end{eqnarray}
where
\begin{equation}
\omega = i \frac{\gamma}{2} \pm \omega_R, \quad \omega_R = \sqrt{k
- \Bigl(\frac{\gamma}{2} \Bigr)^2}.
\end{equation}
On the other hand, the $y$ coordinate describes the amplified motion
\begin{equation}
\ddot{y} - \gamma \dot{y} + k y = 0.
\end{equation}
The amplified solution is
\begin{equation}
y = y_0 e^{i \tilde\omega t},
\end{equation}
where
\begin{equation}
\tilde\omega = - i \frac{\gamma}{2} \pm \omega_R = \omega^*.
\end{equation}

In the quantum theory of the FT oscillator, the Hamiltonian is
given by
\begin{equation}
H_{FT} = p_x p_y + \frac{\gamma}{2} (y p_y - x p_x) + \omega^2_R
xy. \label{dis ham}
\end{equation}
Note that the system keeps the time-reversal symmetry $t
\rightarrow - t$ under $x \leftrightarrow y$. The time-reversal
operation is equivalent to $\omega \leftrightarrow \omega^*$.
Canonical quantization proceeds according to the standard
procedure:
\begin{equation}
[ \hat{x}, \hat{p}_x] = i, \quad [\hat{y}, \hat{p}_y] = i,
\end{equation}
with all other commutators vanishing,
\begin{equation}
[ \hat{x}, \hat{y}] = [\hat{p}_x, \hat{y} ] = [\hat{p}_y, \hat{x}]
= [\hat{p}_x, \hat{p}_y] = 0.
\end{equation}
Using only the real part of the frequency, one may introduce the
annihilation and the creation operators for the system,
\begin{equation}
\hat{a} = \frac{1}{\sqrt{2 \omega_R}} (\hat{p}_x - i \omega_R
\hat{x}), \quad \hat{a}^{\dagger} = \frac{1}{\sqrt{2 \omega_R}}
(\hat{p}_x + i \omega_R \hat{x}),
\end{equation}
and for the environment,
\begin{equation}
\hat{b} = \frac{1}{\sqrt{2 \omega_R}} (\hat{p}_y - i \omega_R
\hat{y}), \quad \hat{b}^{\dagger} = \frac{1}{\sqrt{2 \omega_R}}
(\hat{p}_y + i \omega_R \hat{y}).
\end{equation}
Then, the Hamiltonian has the representation
\begin{equation}
\hat{H}_{FT} = \omega_R ( \hat{a}^{\dagger} \hat{b}^{\dagger} +
\hat{a} \hat{b} ) + i \frac{\gamma}{2} (\hat{a}^{\dagger 2} -
\hat{a}^2 - \hat{b}^{\dagger 2} + \hat{b}^2). \label{ham foc}
\end{equation}
To decouple the coupling between the $x$ and the $y$ coordinates
in the Hamiltonian in Eq. (\ref{ham foc}), one transforms the
annihilation and the creation operators into the new ones as
\begin{equation}
\hat{A} = \frac{1}{\sqrt{2}} (\hat{a} + \hat{b}), \quad \hat{B} =
\frac{1}{\sqrt{2}} (\hat{a} - \hat{b}).
\end{equation}
The transformed operators still satisfy the standard commutation
relations
\begin{equation}
[ \hat{A}, \hat{A}^{\dagger}] = [\hat{B}, \hat{B}^{\dagger}] = 1,
\end{equation}
and all other commutators vanish,
\begin{equation}
[ \hat{A}, \hat{B}] = [\hat{A}, \hat{B}^{\dagger}] =
[\hat{A}^{\dagger}, \hat{B}^{\dagger} ] = 0.
\end{equation}
Finally, one has the Hamiltonian in the new basis
\cite{celeghini}:
\begin{eqnarray}
\hat{H}_{FT} =  \underbrace{\omega_R(\hat{A}^{\dagger} \hat{A} -
\hat{B}^{\dagger} \hat{B} )}_{\hat{H}_0} + \underbrace{i
\frac{\gamma}{2} (\hat{A}^{\dagger} \hat{B}^{\dagger} - \hat{A}
\hat{B})}_{\hat{H}_I}. \label{new ham}
\end{eqnarray}

A few remarks are in order. First, note that the FT oscillator has
the same group structure $SU(1,1)$ as the Hamiltonian of each
quantized quasinormal mode in Eq. (\ref{qnm ham}). In the group
representation obtained from Eq. (\ref{su}) by replacing
$\hat{a}_{\lambda} \rightarrow \hat{A}$ and $\hat{a}_{\bar\lambda}
\rightarrow \hat{B}$, the unperturbed Hamiltonian and the
interaction take the forms
\begin{equation}
\hat{H}_0 = 2 \omega_R \hat{\cal C}, \quad \hat{H}_I = - \gamma
\hat{J}_2.
\end{equation}
Second, the imaginary part, $\omega_{I \lambda}$, of a quasinormal
mode is replaced here by the damping constant $\gamma$. To keep
the correspondence with the FT oscillator, as will be shown below,
the discrete set $\omega_{I n} = (n + 1/2)/4M$ of the imaginary
parts of the quasinormal modes should be identified with the
damping constant as $\gamma = 1/4M$, just one instanton action.
The exponentially damping factor $e^{- (n + 1/2)t/(4M)}$ is a
consequence of quantization.

Soon after Hawking's discovery of pair creation and radiation by
black holes, Israel alternatively derived the black-hole entropy
by using thermofield dynamics (TFD) \cite{israel}. (See also Ref.
\cite{laflamme}.) He observed that in quantizing a field in the
background geometry of a black hole, there is an unobservable
region limited by the horizon in the Kruskal coordinate. This
region plays a role similar to that of the fictitious system of
thermofield dynamics. In thermofield dynamics first introduced by
Takahashi and Umezawa, one doubles the physical system by adding a
fictitious Hamiltonian without any interaction with the system and
uses an extended Hilbert space of the system plus the fictitious
system \cite{takahashi}. The temperature-dependent Bogoliubov
transformation between the annihilation and the creation operators
of the total system yields the thermal state of the physical
system as a two-mode squeezed vacuum state (temperature-dependent
vacuum state) of the extended Hilbert space. Integrating the
quantum field over the unobservable region is equivalent to
tracing over the fictitious system.

The unperturbed Hamiltonian $\hat{H}_0$ in the limit of $\gamma =
0$ may be used to explain the black-hole entropy in line with
Israel's argument. The damped $A$-oscillator describes the
physical system, the exterior of the event horizon of a black
hole, whereas the amplified $B$-oscillator corresponds to a
fictitious system or environment, the interior of the event
horizon. If a temperature-dependent Bogoliubov transformation is
introduced in a suitable way for the total system of $A$- and
$B$-oscillators, the thermal state of the $A$-oscillator is a
vacuum state of the total system. This may correspond to the
static case where one degree of freedom describes a black hole and
the other degree of freedom pertains to the fictitious system.
This black hole entropy would be a kind of entanglement entropy.
The other case of $\gamma \neq 0$ may correspond to the dynamical
situation in which the degree of freedom corresponding to the
black hole interacts continuously with the environmental degree of
freedom. If this correspondence holds true, then the quasinormal
modes may be related with dynamical aspects of black holes.

The $A$-oscillator is even $(A
\leftrightarrow A)$ with respect to the time-reversal symmetry $(x
\leftrightarrow y )$ whereas the $B$-oscillator is odd $(B
\leftrightarrow - B)$. The physical state will be annihilated by
$\hat{B}$ as
\begin{equation}
\hat{B} \vert \psi \rangle_0 = 0.
\end{equation}
In the limit of $\gamma = 0$, $\vert \psi \rangle_0$ is the wave
function of a simple harmonic oscillator. The Hilbert space
is ${\cal H} = {\cal H}_A \otimes
{\cal H}_B$, each subspace of which consists of number states defined as
\begin{eqnarray}
\hat{A}^{\dagger} \hat{A} \vert n_A, n_B \rangle = n_A \vert n_A, n_B \rangle, \nonumber\\
\hat{B}^{\dagger} \hat{B} \vert n_A, n_B \rangle = n_B \vert n_A, n_B \rangle.
\end{eqnarray}
To find the quantum state of the Hamiltonian in Eq. (\ref{new
ham}), one uses two operators that commute with each other, for
instance, $\{ \hat{\cal C}, \hat{J}_3 \}$, and defines their
simultaneous eigenstate as
\begin{eqnarray}
\hat{\cal C} \vert j, m \rangle = j \vert j, m \rangle, \quad j = \frac{1}{2} (n_A - n_B),
\nonumber\\
(\hat{J}_3 - \frac{1}{2}) \vert j, m \rangle = m \vert j, m \rangle,
\quad m = \frac{1}{2} (n_A + n_B). \label{sim eig}
\end{eqnarray}
As the Hamiltonian does not commute with $\hat{J}_3$, one
transforms the eigenstate,
\begin{equation}
\vert \psi_{j, m} \rangle = e^{\frac{\pi}{2} \hat{J}_1} \vert j, m
\rangle,
\end{equation}
then the new state is an eigenstate of the interaction
\begin{equation}
\hat{H}_I \vert \psi_{j, m} \rangle = - i \gamma ( m + \frac{1}{2}) \vert j, m \rangle.
\end{equation}
Note that $- i \gamma ( m + \frac{1}{2})$ is a decaying factor in
real time. As the Casimir operator $\hat{\cal C}$ commutes with
$\hat{J}_1$, the new state is also an eigenstate of the
Hamiltonian in Eq. (\ref{new ham}):
\begin{eqnarray}
\hat{H}_{FT} \vert \psi_{j, m} \rangle = \Bigl[ 2 \omega_R j - i
\gamma ( m + \frac{1}{2}) \Bigr] \vert \psi_{j, m} \rangle.
\end{eqnarray}

A few remarks are in order. The new state $\vert \psi_{j, m}
\rangle$ is not normalizable and diverges, and $\vert \psi_{j, m}
\rangle = e^{\frac{\pi}{2} \hat{J}_1} \vert j, m \rangle$ is not a
unitary transformation in $SU(1,1)$. The dissipative system has a
biorthogonal space \cite{celeghini}. The time-reversal operator
${\cal T}$ is an antiunitary operation:
\begin{eqnarray}
(\hat{A}, \hat{B}) \longrightarrow (- \hat{A}^{\dagger},
- \hat{B}^{\dagger}), \nonumber\\
{\cal T} \vert \psi_{j, m} \rangle = \vert \psi_{j, -(m+1)}
\rangle.
\end{eqnarray}
The dual state is
\begin{eqnarray}
\langle \psi_{j, m} \vert = ({\cal T} \vert \psi_{j,m} \rangle
)^{\dagger},
\end{eqnarray}
such that
\begin{eqnarray}
\langle \psi_{j, m} \vert \psi_{j', m'} \rangle = \delta_{j j'} \delta_{mm'}.
\end{eqnarray}
The solution to the Schr\"{o}dinger equation evolves from an initial state
$\vert j, m_0 \rangle$ as
\begin{eqnarray}
\vert \Psi (t) \rangle = e^{ - 2 i \omega_R j t} \sum_{m \geq |j|}
e^{- \gamma (m+ \frac{1}{2}) t} \langle \psi_{j, m} \vert j, m_0
\rangle \vert \psi_{j, m} \rangle.
\end{eqnarray}
My speculation is that the imaginary parts of quasinormal modes
may have a correspondence with the damping factors of the FT
oscillator as
\begin{eqnarray}
- \frac{i}{4M} (n + \frac{1}{2}) \longleftrightarrow e^{- \gamma
(m + \frac{1}{2}) t},  ~(m, n = 0, 1, 2, \cdots).
\end{eqnarray}
The imaginary parts of quasinormal modes are due to the periodic
motion in Euclidean time, single or multi-instanton actions,
whereas the damping of the FT oscillator is in real time.

\section{Concluding Remarks}

The quasinormal modes of black-hole perturbations are decaying
wave functions that are outgoing to spatial infinity and the event
horizon. Through a study of exactly solvable potentials as
approximations for black-hole potentials, the physical origin of
quasinormal modes is shown to be the single and multi-instantons
of the black-hole potentials. The instanton is a periodic solution
of the inverted potential in Euclidean time and corresponds to a
bound state. The recent proposal that the real parts of
quasinormal-mode frequencies are related with the quantization of
the black-hole area may suggest deep physical implications of
quasinormal modes in black-hole physics. However, the complete
understanding of quasinormal modes should include not only the
real parts but also the imaginary parts of the quasinormal
frequencies. It will be interesting to investigate the origin of
damping factors due to the discrete set of the imaginary
frequencies. As instantons responsible for discrete complex
frequencies are due to the quantization of periodic motions in an
inverted potential, it is likely that a semiclassical approach may
provide better comprehension of quasinormal modes than a classical
approach.

The decaying behavior of quasinormal modes is a characteristic
aspect of dissipative systems. In this paper, I proposed the
quantization of quasinormal modes and a relation with the FT
oscillator, a dissipative open system. The canonical Hamiltonian
for each quasinormal mode has the two-mode representation of the
group $SU(1,1)$. There is an amplified mode in addition to a
damped mode coming from the completeness of mode solutions in
quantizing a field. The imaginary frequency provides an
interaction that connects the damping mode with the amplified
mode. The canonical Hamiltonian of a quasinormal mode has the same
group representation as the FT oscillator. In the FT oscillator,
the damping constant provides the interaction between the damped
and the amplified modes. The energy of the damped mode is
transferred to the amplified mode; thus, the total energy is
conserved. The quantum theory of the FT oscillator, which has been
intensively studied, is expected to give useful information on
black-hole physics via quasinormal modes. The quantum theory of
the FT oscillator and/or the quasinormal mode of a black hole is
nonunitary because it is a dissipative open system. An interesting
model in this direction is provided by three-dimensional BTZ black
holes which allow exact solutions and have AdS/CFT correspondence
\cite{myung}. It is shown there that a nonrotating BTZ black hole
having quasinormal modes is nonunitary and, thus, belongs to a
dissipative system. This nonunitarity is in agreement with
conformal field theory \cite{cardoso,birmingham}. On the other
hand, an extreme BTZ black hole or pure AdS spacetime has only
real quasinormal-mode frequencies and, thus, is unitary. This
implies that quasinormal modes are related with a dissipative
system and nonunitarity.

Finally, it would be very interesting to find a connection between
quasinormal mode and black-hole thermodynamics, if any. From the
arguments in this paper, that is highly likely. This point is
under study.

{\it Note added}. After submission of this paper, I was informed
by several authors of many relevant references. Berti {\it et al}
pointed out in Refs. \cite{berti1,berti2} that the real parts of
the highly damped quasinormal modes of the Kerr black hole do {\it
not} have the asymptotic frequencies suggested by Hod. This fact,
however, does not significantly change the arguments of this paper
because the imaginary parts of the quasinormal modes are related
with dissipative open systems in the quantization of the
quasinormal modes. Padmanabhan {\it et al} used the first Born
approximation to calculate the scattering matrix of the
quasinormal modes, from which the discrete complex frequencies are
derived \cite{padmanabhan1,padmanabhan2}. Setare {\it et al} also
discussed the black-hole area quantization for a non-rotating BTZ
black hole, an extremal Reissner-Nordstr\"{o}m black hole, and an
extremal Kerr black hole \cite{setare1,setare2,setare3}.

\acknowledgements

The author would like to thank Valeri Frolov, Viqar Husain, Werner
Israel, Gungwon Kang, Gabor Kunstatter, and Don N. Page for useful
discussions and information. This work was supported by the Korea
Science and Engineering Foundation under R01-2005-000-10404-0.


\begin{references}

\bibitem{vishveshwara} C. V. Vishveshwara, Nature {\bf 227}, 396 (1970).

\bibitem{hod} S. Hod, Phys. Rev. Lett. {\bf 81}, 4293 (1998).

\bibitem{horowitz} G. T. Horowitz and V. E. Hubeny, Phys. Rev. D
{\bf 62}, 024027 (2000).

\bibitem{kunstatter} G. Kunstatter, Phys. Rev. Lett. {\bf 90},
16301 (2003).

\bibitem{dreyer} O. Dreyer, Phys. Rev. Lett. {\bf 90}, 081301
(2003).

\bibitem{kokkotas} K. D. Kokkotas and B. G. Schmidt, Living Rev. Rel.
{\bf 2}, 2 (1999).

\bibitem{natario} J. Nat\'{a}rio and R. Schiappa, ``On the Classification of
Asymptotic Quasinormal Frequencies for d-Dimensional Black Holes
and Quantum Gravity,'' hep-th/0411267 (2004).

\bibitem{kim0} S. P. Kim, Bull. Korean Phys. Soc. {\bf 20},  2, 565
(2002).

\bibitem{kim1} S. P. Kim, ``Quasinormal Modes of Black Holes and
Dissipative Open Systems,'' talk at Black Hole V (2005)
(unpublished).

\bibitem{bekenstein} J. D. Bekenstein, Lett. Nuovo Cimento {\bf
11}, 467 (1974).

\bibitem{nollert} H. P. Nollert, Phys. Rev. D {\bf 47}, 5253
(1993).

\bibitem{hod2} S. Hod, Phys. Rev. D {\bf 67}, 081501 (2003).

\bibitem{barton} G. Barton, Ann. Phys. {\bf 166}, 322 (1986).

\bibitem{kim2} S. P. Kim and D. N. Page, Phys. Rev. D {\bf 65}, 105002 (2002).

\bibitem{roberts} M. D. Roberts, Gen. Relativ. Gravit. {\bf 21},
907 (1989).

\bibitem{bak1} D. Bak, S. K. Kim, S. P. Kim, K-S. Soh, and J. H. Yee, Phys.
Rev. D {\bf 60}, 064005 (1999).

\bibitem{bak2} D. Bak, S. K. Kim, S. P. Kim, K-S. Soh, and J. H. Yee, Phys.
Rev. D {\bf 61}, 044005 (2000).

\bibitem{bak3} D. Bak, S. K. Kim, S. P. Kim, K-S. Soh, and J. H. Yee, Phys.
Rev. D {\bf 62}, 047504 (2000).

\bibitem{ferrai} V. Ferrai and B. Mashoon, Phys. Rev. Lett. {\bf
52}, 1361 (1984).

\bibitem{ferrai2} V. Ferrai and B. Mashoon, Phys. Rev. D {\bf 30}, 295 (1984).

\bibitem{motl} L. Motl, Adv. Theor. Math. Phys. {\bf 6}, 1135 (2003).

\bibitem{motl2} L. Motl and A. Neitzke, Adv. Theor. Math. Phys. {\bf 7}, 307 (2003).

\bibitem{schroer} B. Schroer and J. A. Swieca, Phys. Rev. D {\bf 2}, 2938 (1970).

\bibitem{schroer2} B. Schroer, Phys. Rev. D {\bf 3}, 1764 (1971).

\bibitem{matacz} A. L. Matacz, P. C. W. Davies, and A. C. Ottewill, Phys. Rev. D
{\bf 47}, 1557 (1993).

\bibitem{kang} G. Kang, Phys. Rev. D {\bf 55}, 7563 (1997).

\bibitem{mukohyama} S. Mukohyama, Phys. Rev. D {\bf 61}, 124021 (2000).

\bibitem{fulling} S. A. Fulling, {\it Aspects of Quantum Field
Theory in Curved Space-Time} (Cambridge Univ. Press, Cambridge,
1989).

\bibitem{hawking} S. W. Hawking, Comm. Math. Phys. {\bf 43}, 199
(1975).

\bibitem{beyer} H. R. Beyer, Commun. Math. Phys. {\bf 204}, 397 (1999).

\bibitem{price} R. H. Price and V. Husain, Phys. Rev. Lett. {\bf 68}, 1973.

\bibitem{ching} E. S. C. Ching, P. T. Leung, W. M. Suen, and K.
Young, Phys. Rev. D {\bf 54}, 3778 (1996).

\bibitem{ho} K. C. Ho, P. T. Leung, A. M. van den Brink, and K.
Young, Phys. Rev. E {\bf 58}, 2965 (1998).

\bibitem{feshbach} H. Feshbach and Y. Tikochinsky,
Trans. N.Y. Acad. Sci. {\bf 38} (Ser. II), 44 (1977).

\bibitem{kim3} S. P. Kim, A. E. Santana, and F. C. Khanna, J.
Korean Phys. Soc. {\bf 43}, 452 (2003).

\bibitem{kim4} S. P. Kim, J. Korean Phys. Soc. {\bf 44}, 446
(2004).

\bibitem{celeghini} E. Celeghini, M. Rasetti, and G. Vitiello,
Ann. Phys. {\bf 215}, 156 (1992).

\bibitem{israel} W. Israel, Phys. Lett. A {\bf 57}, 107 (1976).

\bibitem{laflamme} R. Laflamme, Nucl. Phys. B {\bf 324}, 233
(1989).

\bibitem{takahashi} Y. Takahashi and H. Umezawa, Collective
Phenomena {\bf 2}, 55 (1975) [reprinted in Int. J. Mod. Phys. B
{\bf 10}, 1755 (1996)].

\bibitem{myung} Y. S. Myung and H. W. Lee, ``Unitarity Issue in BTZ Black
Holes,'' hep-th/0506031.

\bibitem{cardoso} V. Cardoso and J. P. S. Lemos,
Phys. Rev. D {\bf 63}, 124015 (2001).

\bibitem{birmingham} D. Birmingham, I. Sachs, and S. Solodukhin,
Phys. Rev. Lett. {\bf 88}, 151301 (2002).

\bibitem{berti1} E. Berti, V. Cardoso, K. D. Kokkotas, and H. Onozawa,
Phys. Rev. D {\bf 68}, 124018 (2003).

\bibitem{berti2} E. Berti, V. Cardoso, and S. Yoshida,
 Phys. Rev. D {\bf 69}, 124018 (2004).

 \bibitem{padmanabhan1} T. Padmanabhan,
 Class. Quantum Grav. {\bf 21}, L1 (2004).

\bibitem{padmanabhan2} T. R. Choudhury and T. Padmanabhan,
Phys. Rev. D D {\bf 69}, 064033 (2004).

\bibitem{setare1} M. R. Setare, Class. Quantum Grav. {\bf 21},
1453 (2004).

\bibitem{setare2} M. R. Setare, Phys. Rev. D {\bf 69}, 044016
(2004).

\bibitem{setare3} M. R. Setare and E. C. Vagenas,
Mod. Phys. Lett. A {\bf 20}, 1923 (2005).





\end{references}
\end{document}